\documentclass[aps,12pt,final,notitlepage,oneside,onecolumn,nobibnotes,nofootinbib,
superscriptaddress,noshowpacs,centertags]{revtex4-1}
\usepackage[utf8]{inputenc}
\usepackage[english]{babel}
\usepackage{amsmath}
\usepackage{amsfonts}
\usepackage{amssymb}
\usepackage{graphicx}
\def\bra{\langle}
\def\ket{\rangle}

\begin{document}
\title{Symmetry and decoherence-free subspaces in quantum neural networks}
\author{Altaisky M.V.}
\email{altaisky@rssi.ru}
\affiliation{Space Research Institute RAS, Profsoyuznaya 84/32, Moscow, 117997, Russia}

\author{N. E. Kaputkina}
\affiliation{National Technological University ''MISIS'', Leninsky prospect 4, Moscow, 119049, Russia}
\email{nataly@misis.ru}

\author{V. A. Krylov}
\affiliation{Joint Institute for Nuclear Research, Joliot Curie 6, Dubna, 141980, Russia}
\email{kryman@jinr.ru}

\begin{abstract}
Evolution of quantum states of array of quantum dots is analyzed by means of numerical solution of the von
Neumann equation. For two qubit system with dipole-dipole interaction and common phonon bath
the evolution of the symmetric state $\frac{\uparrow\downarrow+\downarrow\uparrow}{\sqrt{2}}$ leads to the 
mixture of the triplet states, leaving 
the singlet  decoupled. For three qubit system ($D_{1/2}^{\otimes3}=D_{3/2}+2D_{1/2}$) with  
common phonon bath we observed similar effects within the quartet 
state $D_{3/2}$ if  all qubits were symmetrically connected. 
\end{abstract}

\maketitle
\section{Introduction}
Both the increasing data flows of modern computations and 
the miniaturization of data processing units imply  the future high performance computers will be based on some quantum elements, which obey the 
laws of quantum mechanics. Quantum computational devices, being first proposed by R.Feynman \cite{Feynman1982}, 
can solve the exponentially hard problems of classical computing 
in polynomial time \cite{DJ1992}. The main obstacle on the way to a scalable  quantum computer is the decoherence induced by interaction with environment. This keeps the 
range of commercially affordable circuit-based computers within a few qubits. The compromise between decoherence and 
quantum parallelism have been achieved rather recently in a form of adiabatic quantum computers -- a quantum neural networks made of SQUIDS \cite{D-wave2011s}. This is not a quantum computer in a strict sense of \cite{DJ1992}, but a quantum simulator which mimics the work of the Hopfield 
network using quantized magnetic fluxes of SQUIDS. Being restricted in the range of problems to be solved, the quantum neural network capacity 
dramatically increases due to quantum parallelism.

Quantum neural network is a massively 
parallel processing unit made of a large set of identical quantum elements, interacting to each other by many possible, but controlled ways, capable 
of a self-organization, called {\em learning} \cite{Haykin1999}.
The set of elements may be distinguishable, and hence, directly addressable,
as in SQUID network \cite{D-wave2011s}, or may be indistinguishable (like the atoms of Bose gas), but still interacting to each other in a controlled way \cite{berloff2017}. The learning and 
the performance rates of a massively parallel quantum processing unit 
thus depend on its symmetry properties.  

\section{Quantum Neural Networks}
In contrast to  visible progress in circuit-based quantum computing \cite{DJ1992}, the quantum neural network studies were out  
from the mainstream of quantum information processing until the first hardware implementation of the quantum Hopfield network has been built by D-Wave systems Inc. \cite{D-wave2011}. The adiabatic 
quantum computers produced by D-wave systems Inc. are described by the Hamiltonian of the 
form
\begin{equation}
H = \sum_i K_i \sigma^x_i +\sum_i H_i \sigma^z_i + \sum_{i\ne j}J_{ij} \sigma^z_i\sigma^z_j, \label{Hdw}
\end{equation}
with the ''spins'' implemented by SQUID elements with two possible directions of magnetic flux, and the problem matrix $J_{ij}$, implemented as a set of inductive couplings between the SQUIDs. The general idea beneath adiabatic quantum computing 
is to evolve adiabatically the system of identical quantum elements, described 
by the Hamiltonian of the form 
\begin{equation} \label{aqc}
H(t) = \Gamma(t) H_F + \Lambda(t) H_P, \quad \Gamma(0)=1,  \Gamma(\infty)=0,
 \Lambda(0)=0,  \Lambda(\infty) = 1,                   
\end{equation} 
where at initial time $t=0$ the system is hold in the ground state 
of fiducial Hamiltonian $H_F$, to the ground 
state of the problem Hamiltonian $H_P$ at $t\to\infty$. 

Referring the reader to the original papers on adiabatic quantum computers on SQUIDs \cite{D-wave2011,CT2014,Schuld2014}, we would like to emphasize their shortage is a very low operational temperature  of about $10^{-1}$K range. This results in a high energy consumption of the cooling system and prevents the construction 
of portable devices. There is a quest for alternative elements 
for quantum neural networks with the operational temperature higher than that of SQUIDs. (Needless to say that the brain 
network itself may be a quantum neural network \cite{Chav1970e,BE1992}.)
One of the alternatives promising a quantum neural network operating at 
room temperature may be quantum dot based quantum neural networks \cite{BN2000,AKK2014,Alt2016apl}. The advantage of quantum dots as elements 
of quantum neural networks is the flexibility of control of their characteristics, i.e, energy spectra, electron localization radius, interaction strength, by external magnetic field \cite{KL1998pss,Kaputkina1998}.

Regardless particular type of quantum elements, considered as quantum neurons, the obedience of such system to the laws of quantum mechanics 
assumes the importance of the symmetries of the interaction Hamiltonian. The 
symmetries affect both the energy spectra and the structure of space of states of quantum system. The individual 
elements of the network cannot be directly addressed, but only the collective states of the whole network can be controlled.

Hopfield network can either work as minimizer, that search  for an optimal configuration 
$(s_0,\ldots,s_N),s_i=\pm1$, that minimize the energy 
$E(s)= \sum h_i s_i + \sum_{i\ne j} J_{ij}s_i s_j$, or as an associative memory, that seeks for an optimal matrix $J_{ij}$ to project an arbitrary input vector $(s_0,\ldots,s_N)$ into one of $M\le N$ possible 
attractors. For quantum case, specially when all qubits are implemented on 
SQUIDs or quantum dots, it is impossible to change the elements of the 
connection matrix continually, say by gradient descent method. In contrast, 
we can set either $J_{ij}=1$, for connected, or $J_{ij}=0$ for disconnected pairs \cite{D-wave2011}. 
The problem of quantum neural network is a discrete optimization 
problem with respect to either ''spins'' $s_i$, or connection matrix $J_{ij}$, 
both affected by environment, i.e., the strength of external driving 
field $K_i$, the temperature of the phonon bath, the spectral density of phonons, etc. Driving such system from a specially initial quantum state 
to the final state with minimal energy is a typical problem for Boltzmann 
machines. It can be solved by simulated annealing in classical case, or 
by quantum annealing in quantum case.

In the remaining part of this paper we will study the role of symmetries for a toy model of three-qubit network with dipole-dipole interaction  and common phonon bath.     

\section{Three qubit system dynamics}
Our toy model consists of $N=3$ identical qubits with two orthogonal quantum states, $|X\ket$ for eXcited, and $|0\ket$ for ground state, interacting to each other by means of dipole-dipole coupling, and linearly interacting to 
the common heat bath. It is described by the Hamiltonian:  
\begin{equation}\nonumber 
H =\sum_{i=0}^{N-1} \frac{\delta_i}{2} (\sigma_z^{(i)}+1) + \sum_{i=0}^{N-1} \frac{K_i}{2} \sigma_x^{(i)} 
+ \sum_{i\ne j} J_{ij}\sigma_+^{(i)}\sigma_-^{(j)}  
+ \sum_{a,i} g_a x_a |X_i\ket\bra X_i| + H_{Ph} \equiv H_0+H_{Int}+H_{Ph},
\label{Hamiltonian3SRW}
\end{equation}
written in rotating wave approximation, with $\delta_i$ being the detuning of the driving electric field frequency from the $i$-th qubit resonance frequency $\frac{E^ i_X-E^i_0}{\hbar}$; 
 $\hbar=1$ assumed hereafter;  $K_i$ is a coupling to an external driving field, $J_{ij}$ is the dipole-dipole coupling between qubits. The phonon modes $x_a$ are assumed to interact only to the excited states $|X_i\ket$ \cite{RQJ2002}.
The pseudo-spin operators of the $i^\textrm{th}$ qubit are: 
\begin{align*}
\sigma^{(i)}_z = |\textrm{X}_i\ket\bra \textrm{X}_i| - |0_i\ket\bra 0_i|,\quad  
\sigma^{(i)}_x = |0_i\ket\bra \textrm{X}_i| + |\textrm{X}_i\ket\bra 0_i|, \quad 
\sigma^{(i)}_+ = |\textrm{X}_i\ket\bra 0_i|,\quad   
\sigma^{(i)}_- = |0_i\ket\bra \textrm{X}_i|.
\end{align*}
Our toy model is an extension of quantum dot interaction model  considered in \cite{Alt2016apl}.

\begin{figure}[ht]
\centering \includegraphics[width=6cm]{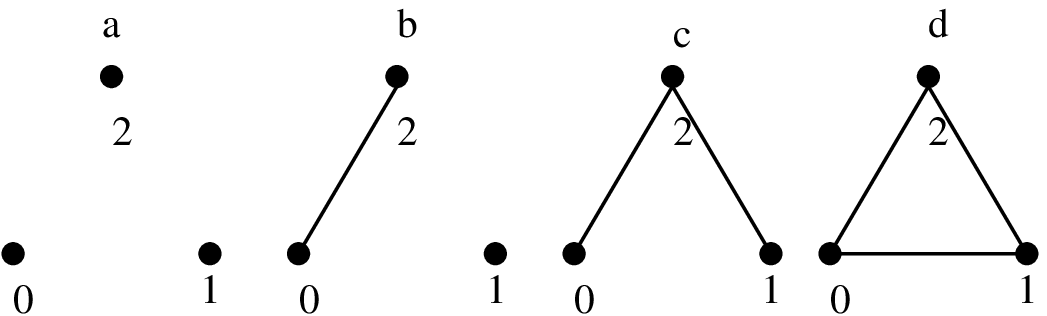}
\caption{Possible connections in three qubit network}
\label{top3:pic}
\end{figure}

\begin{figure}[ht]
\centering \includegraphics[width=7cm]{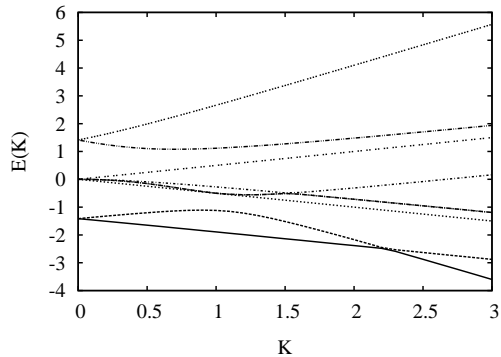}
\caption{Energy levels of $\Lambda$-type configuration. $E(K)$ are plotted in $J\!=\!1$ units}
\label{h20:pic}
\end{figure}

In case of 3 qubits there are 8 classically distinguished states, from $000$ to $XXX$, and also 8 topologically 
distinguishable types of connection between qubits, $a+3b+3c+d$, see 
Fig.~\ref{top3:pic}.
In case of indistinguishable qubits there are only 4 such connection, corresponding to the cases a,b,c,d, respectively.
Case 'a' is a trivial case of no network, case 'b' corresponds to two qubits considered in our paper \cite{Altaisky201724}, so only 'c' and 'd' can be considered as the connections of 3 qubit network.
At the absence of interaction with environment, the energy levels of such systems can be easily evaluated.  
In standard computational basis ($000,00X,\ldots,XXX$) the free part of the 
Hamiltonian $H$ can be written as 
\begin{equation}
H_0 = \begin{pmatrix}
0 & \frac{K_0}{2} & \frac{K_1}{2} & 0 & \frac{K_2}{2} & 0 & 0 & 0 \cr
\frac{K_0}{2} & 0 & J_{01} & \frac{K_1}{2} & J_{02} & \frac{K_2}{2} & 0 & 0 \cr
\frac{K_1}{2} & J_{01} & 0 & \frac{K_0}{2} & J_{12} & 0 & \frac{K_2}{2} & 0 \cr
0 & \frac{K_1}{2} & \frac{K_0}{2} & 0 & 0 & J_{12} & J_{02} & \frac{K_2}{2} \cr
\frac{K_2}{2}& J_{02} & J_{12} & 0 & 0 & \frac{K_0}{2} & \frac{K_1}{2} & 0 \cr
0 & \frac{K_2}{2} & 0 & J_{12} & \frac{K_0}{2} & 0 & J_{01} & \frac{K_1}{2} \cr 
0 & 0 & \frac{K_2}{2} & J_{02} & \frac{K_1}{2} & J_{01} &0 &
\frac{K_0}{2} \cr
0&0&0&\frac{K_2}{2}&0&\frac{K_1}{2}& \frac{K_0}{2}& 0
 
\end{pmatrix}\label{H03}
\end{equation}
with all detunings set to zero $\delta_i=0$. Let us start with the symmetric case 'd'. Setting in \eqref{H03} all connections equal $J_{02}=J_{01}=J_{12}=J$, and a common driving field $K_0=K_1=K_2=K$
We get the energy eigenvalues 
$$
E_{1,2} = J + \frac{K}{2} \pm \sqrt{K^2+JK+J^2}, \quad
E_{3,4} = J - \frac{K}{2} \pm \sqrt{K^2-JK+J^2}, \quad 
E_{5,6,7,8} =-J \pm \frac{K}{2}.
$$
The last two eigenvalues are doubly degenerated. 
\begin{figure}[ht]
\centering \includegraphics[width=7cm]{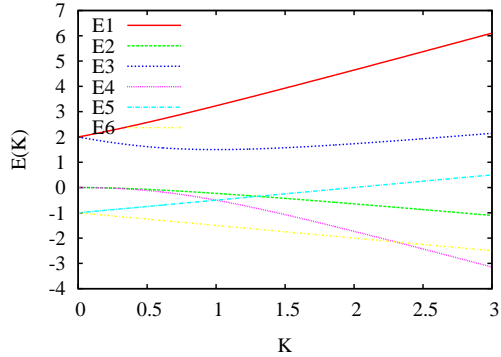}
\caption{Energy levels of symmetric $\triangle$-type configuration Hamiltonian $H_0$. $E(K)$ are plotted in $J\!=\!1$ units}
\label{E1-6:pic}
\end{figure}
For less symmetric configuration of $\Lambda$-shape, shown in Fig.~\ref{top3:pic}'c',  the energy gap 
between low-lying negative energy states and the excited states can 
be effectively suppressed by increasing the driving field strength $K$.
The energy spectrum of the Hamiltonian $H_0$ with $J_{01}=J_{02}=J,J_{12}=0$ 
is shown in Fig.~\ref{h20:pic}. 
If the system of coupled qubits interacts to a heat bath of phonons, an 
initial quantum state of such system, prepared as a linear superposition of 
quantum states of computational basis, will decohere into a mixture of quantum states, unless it does not belong to a {\em decoherence-free subspace}, i.e., is not subjected to decoherence \cite{ZR1997}.

The evolution of 3 qubit quantum system is described by the von Neumann equation 
for the density matrix 
\begin{equation}
\dot{\rho} = - \frac{\imath}{\hbar}[H,\rho],
\label{vn}
\end{equation}  
where $\rho(t)$ is the total density matrix of the system together with 
its environment, with the initial state decoupling condition 
$
\rho(0) = \rho_{SYS}(0) \otimes \rho_{ENV}(0).
$
The interaction between the system and environment is assumed be weak 
enough to write the total Hamiltonian in the form $
H = H_{SYS} + H_{ENV} + \lambda H_{INT}$, 
where $\lambda$ is a small parameter. The interaction Hamiltonian
$H_{INT}$ is usually assumed to be linear in system coordinates. 
From the  theory of open quantum system \cite{FV1963}, we 
expect that the fluctuations of environment play the same role in 
quantum neural networks as they do in any Boltzmann machine, and 
the system density matrix $\rho_{SYS}(t)$ will evolve to the minimal energy 
state of the system. The evolution equation for the system density matrix is 
given by tracing over the environment degrees of freedom 
\begin{equation}
\dot{\rho}_{SYS} = - \frac{\imath}{\hbar}{\rm Tr}_{ENV}[H,\rho],
\label{vns}
\end{equation} 
where $\rho_{SYS}= {\rm Tr}_{ENV}\rho$. The latter equation \eqref{vns} can 
be solved either perturbatively, or numerically, using the numerical 
methods for quasi-adiabatic path integral \citep{1995Makri2}.

The quantitative measures of the ''quantumness'' of collective states of 
quantum system are their entanglement and different types of entropy \cite{BVSW1996}. 

Following our previous papers studying two dipole-dipole coupled qubits with common phonon bath \cite{AZ2016epj,Altaisky201724}, where we have observed that initially symmetric state 
$$
|e_3\ket = \frac{|X0\ket+|0X\ket }{\sqrt{2}}
$$
is spread into a uniform mixture of three triplet states (in the Bell basis) 
\begin{equation}
|e_1\ket = \frac{|XX\ket+|00\ket }{\sqrt{2}}, 
|e_2\ket = \frac{|XX\ket-|00\ket}{\sqrt{2}},
|e_3\ket,
\end{equation}
without mixing to a singlet state 
$
|e_4\ket = \frac{|X0\ket-|0X\ket }{\sqrt{2}}
$   
even in the presence of phonon bath.

This separation to be a consequence of the decomposition 
of direct product of two spin Hilbert spaces into a sum of 
invariant subspaces 
$
D_{\frac{1}{2}} \otimes D_{\frac{1}{2}} = \underbrace{D_1}_{triplet} \oplus \underbrace{D_0}_{singlet}.
$

Based on the $SU(2)$ representation structure of the three spin qubits states  
\begin{equation}
D_{\frac{1}{2}}\otimes D_{\frac{1}{2}}\otimes D_{\frac{1}{2}} =D_{\frac{3}{2}}\oplus 2D_{\frac{1}{2}},
\end{equation}
where the quartet representation $D_{\frac{3}{2}}$ is totally symmetric 
with respect to all three spins, we can conjecture similar structure 
of invariant subspaces, surviving at the presence of phonon bath, 
at least for symmetric coupling between qubits. 

Using the standard computational basis 
$(e_0 = |000\ket, \ldots, e_7 = |XXX\ket)$ we can construct 
the basis for quartet representation $D_{3/2}$:
\begin{align*}
f_{\frac{3}{2},+\frac{3}{2}} &=& |XXX\ket &=& e_7 , \\
f_{\frac{3}{2},+\frac{1}{2}}    &=& \frac{ |0XX\ket + |X0X\ket + |XX0\ket}{\sqrt{3}} &=& 
\frac{ e_3 + e_5 + e_6}{\sqrt{3}}, \\
f_{\frac{3}{2},-\frac{1}{2}}    &=& \frac{|X00\ket + |0X0\ket + |00X\ket }{\sqrt{3}} &=& 
\frac{e_4 + e_2 + e_1}{\sqrt{3}}, \\
f_{\frac{3}{2},-\frac{3}{2}} &=& |000\ket &=& e_0.
\end{align*}
The bases for two remaining $D_{1/2}$ representations can be 
constructed as the set of vectors orthogonal to the $f_{\frac{3}{2},+\frac{1}{2}}$ and $f_{\frac{3}{2},-\frac{1}{2}}$:
\begin{align*}
f_4 = - \sqrt{\frac{2}{3}} e_3 + \frac{1}{\sqrt{6}} e_5 + \frac{1}{\sqrt{6}} e_6, & &
f_6 = - \sqrt{\frac{2}{3}} e_4 + \frac{1}{\sqrt{6}} e_2 + \frac{1}{\sqrt{6}} e_1,
\\
f_5 = \frac{1}{\sqrt{2}}e_5 - \frac{1}{\sqrt{2}}e_6,& & 
f_7 = \frac{1}{\sqrt{2}}e_2 - \frac{1}{\sqrt{2}}e_1.
\end{align*}
They form two $D_{1/2}$ doublets: $(f_5,f_7)$ and $(f_4,f_6)$.
Similar to the Bell basis, the states $e_0$ and $e_7$ can be symmetrized 
$
f_0 = \frac{e_0+e_7}{\sqrt{2}}, \quad f_1 = \frac{e_0-e_7}{\sqrt{2}}.
$
In matrix form the group basis $f_k$ is related to the computational 
basis $e_l$ : $f_k = A_{kl}e_l,$ where 
\begin{equation}
A = \begin{pmatrix}
\frac{1}{\sqrt{2}} & 0 & 0 & 0 & 0 & 0 & 0 &  \frac{1}{\sqrt{2}} \cr
\frac{1}{\sqrt{2}} & 0 & 0 & 0 & 0 & 0 & 0 & -\frac{1}{\sqrt{2}} \cr
0 & 0 & 0 & \frac{1}{\sqrt{3}} & 0 &\frac{1}{\sqrt{3}}& \frac{1}{\sqrt{3}}& 0 \cr
0 & \frac{1}{\sqrt{3}}& \frac{1}{\sqrt{3}}& 0 &\frac{1}{\sqrt{3}} & 0 & 0 & 0 \cr
0&0&0& - \frac{2}{\sqrt{6}} & 0 & \frac{1}{\sqrt{6}} & \frac{1}{\sqrt{6}} & 0 \cr 
0&0&0&0&0& \frac{1}{\sqrt{2}} & -\frac{1}{\sqrt{2}} & 0 \cr
0 & \frac{1}{\sqrt{6}} & \frac{1}{\sqrt{6}} & 0 & - \frac{2}{\sqrt{6}} & 
0&0&0 \cr
0 &  -\frac{1}{\sqrt{2}} & \frac{1}{\sqrt{2}} & 0 &0 &0 &0 &0 
 
\end{pmatrix}
\end{equation} 
We performed numeric simulations with symmetric configuration with density matrix 
$\hat{\rho}(0) = |f(3/2,+1/2)\ket \bra f(3/2,+1/2)|$ 
and calculate the evolution of density matrix for the super-ohmic 
spectral density of phonon bath, and $T=77K$. 
\begin{figure}[ht]
\centering \includegraphics[width=8cm]{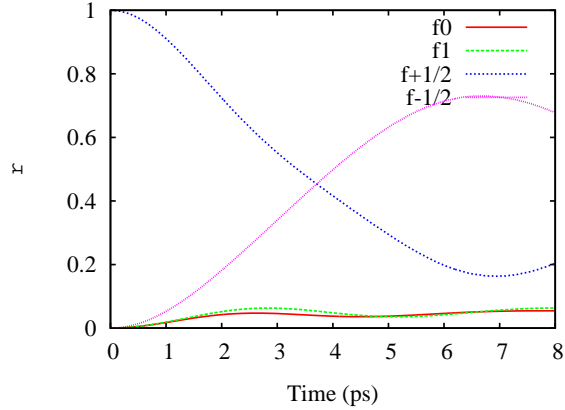}
\caption{Evolution of diagonal density matrix elements in group 
basis for $D_{\frac{3}{2}}$ with $\triangle$-type Hamiltonian evaluated by quasi-adiabatic path integral method. The diagonal elements of the orthogonal spaces $D_{1/2}$ a kept zero.
The nondiagonal elements are not shown.}
\label{vn32s:pic}
\end{figure}
In numerical simulations the values of the coupling constants $K=0.24,J=0.595$, in accordance to quantum dot sizes and parameters given in \cite{AZ2016epj}. The results of simulations are presented at the phonon bath temperature $T=77K$.

If the interaction Hamiltonian is not symmetric to permutations the initial 
symmetric state will span the whole basis during the time evolution, as is 
displayed in Fig.~\ref{vn32l:pic} below.
\begin{figure}[ht]
\centering \includegraphics[width=7cm,angle=270]{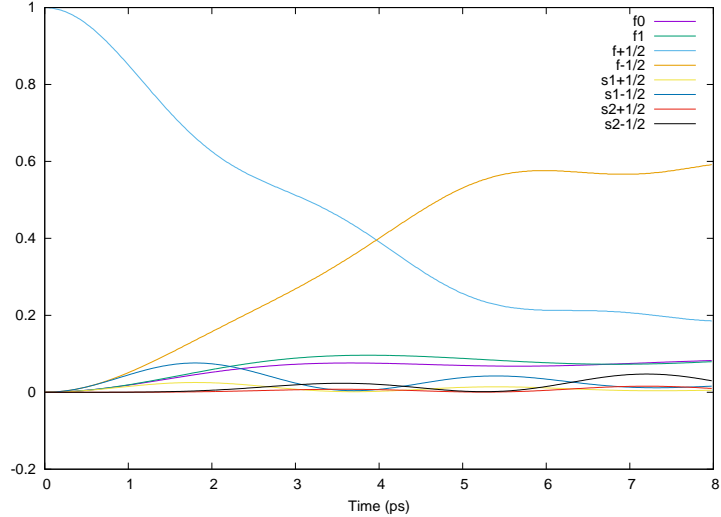} 
\caption{Evolution of diagonal density matrix elements in group 
basis for $D_{\frac{3}{2}} + 2D_\frac{1}{2}$ for $\Lambda$ type Hamiltonian and 
symmetric $f(\frac{3}{2},+\frac{1}{2})$ initial state}
\label{vn32l:pic}
\end{figure}
Starting from initial state $\rho(0)=|f_7\ket \bra f_7|$ we have the oscillations between $f_7$ and $f_5$
\begin{figure}[ht]
\centering \includegraphics[width=7cm,angle=270]{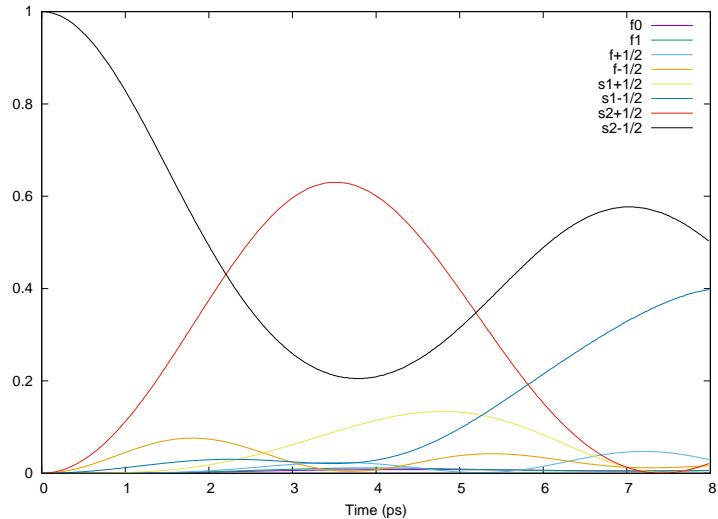}   
\caption{Evolution of diagonal density matrix elements in group 
basis for $D_{\frac{3}{2}} + 2D_\frac{1}{2}$ for $\triangle$ type Hamiltonian and $f_7$ initial state}
\label{f5f7:pic}
\end{figure}

\section{Conclusion}
We conclude that the ideas of the decoherence-free subspaces, proposed for singlet states \cite{ZR1997}, can be exploited for rather general open 
quantum systems of  quantum dots. Depending on 
the symmetry of connection between the dots, the states of the  
array can be hold within particular multiplets, the sum of which 
comprise the space of quantum states of the array. This constrains 
environment-induced decoherence  and provides possibilities for building adiabatic quantum computers working 
at high temperatures.
%

\end{document}